# Rejoinder: Gibbs Sampling, Exponential Families and Orthogonal Polynomials

**Persi Diaconis, Kshitij Khare and Laurent Saloff-Coste**

*Abstract.* We are thankful to the discussants for their hard, interesting work. The main purpose of our paper was to give reasonably sharp rates of convergence for some simple examples of the Gibbs sampler. We chose examples from expository accounts where direct use of available techniques gave practically useless answers. Careful treatment of these simple examples grew into bivariate modeling and Lancaster families. Since bounding rates of convergence is our primary focus, let us begin there.

## 1. RATES OF TWO COMPONENT GIBBS SAMPLERS

Consider the beta/binomial example (with a uniform prior) discussed in our introduction. Some of our students tried to use the Harris recurrence techniques directly on the two component chain. The two-component chain $\widetilde{K}$ goes $(x, \theta) \to (x, \theta') \to (x', \theta')$. To establish the drift condition: $E(V(X_{i+1}) \mid X_i = x) \leq \lambda V(x) + b$, they chose $V(x, \theta) = x$. Then $E(x' \mid (x, \theta)) = \frac{nx}{n+2} + \frac{n}{n+2}$, so $\lambda \geq \frac{n}{n+2}, b = \frac{n}{n+2}$ work. For the minorization condition, they used the factorization

$$f((x', \theta') \mid (x, \theta)) = f_1(\theta' \mid x) f_2(x' \mid \theta')$$

with $f_1(\cdot \mid x)$ the Beta$(x+1, n-x+1)$ density and $f_2(\cdot \mid \theta')$ the Binomial$(n, \theta')$ density. Let $g(\theta) =$ $\inf_x f_1(\theta \mid x)$, $\epsilon = \int_0^1 g(\theta) \, d\theta = \frac{1}{2^n}$ and $q(x, \theta) = \epsilon^{-1} g(\theta) f_2(x \mid \theta)$. Then the minorization condition

$$f((x', \theta') \mid (x, \theta)) \geq \epsilon q(x, \theta) \quad \forall x, x', \theta, \theta'$$

is satisfied. This leads to the bound

$$\|\widetilde{K}_{x_0}^\ell - f\|_{\mathrm{TV}}$$
$$\leq (1-\epsilon)^{r\ell} + \left(\frac{u^r}{\alpha^{1-r}}\right)^\ell \left(1 + \frac{b}{1-\lambda} + V(x_0)\right),$$

with

$$\alpha = \frac{1+d}{1+2b+\lambda d},$$
$$u = 1 + 2(\lambda d + b), \quad d \geq \frac{2b}{1-\lambda}.$$

For $n = 100$ and $x_0 \in (0, \frac{1}{2})$, they chose $d = 1000$, $r = \frac{1}{1000}$. The bound says that if we run the sampler $10^{33}$ steps, the total variation distance will be less than 0.01.

A similarly poor rate follows from Proposition and Example 4.1.1 of Berti et al. The point of spelling out this example is *not* to make fun of anyone, but to emphasize how a reasonable first pass at using off the shelf tools can lead to a useless answer. Here, despite the fact that an explicit eigenfunction corresponding to the largest eigenvalue was available as a choice for the drift function $V$.

We are impressed and thankful to Berti et al. and Jones and Johnson for carrying out the work to massage their bounds into a more useable form. We regard the treatment of the normal example in

*Persi Diaconis is Professor, Department of Statistics, Sequoia Hall, 390 Serra Mall, Stanford University, Stanford, California 94305-4065, USA e-mail: diaconis@math.stanford.edu. Kshitij Khare is Graduate Student, Department of Statistics, Sequoia Hall, 390 Serra Mall, Stanford University, Stanford, California 94305-4065, USA e-mail: kdkhare@stanford.edu. Laurent Saloff-Coste is Professor, Department of Mathematics, Cornell University, Malott Hall, Ithaca, New York 14853-4201, USA e-mail: lsc@math.cornell.edu.*

Jones and Johnson as particularly successful (we don't see any practical difference between "3 steps" and "99 steps"). The reader who studies their argument will find clever, nonobvious choices coupled with computer work. Proposition 3 in Berti et al. also seems quite useful. They say they are playing "devils advocate." We note that "the devil is in the details." We have tried their suggestion to use "numerical evaluations" to make a choice of $d, r$ and $B$ from their Proposition 3 for their example 4.2.1. After some playing around, the best we found is $d = 4, r = 0.11, B = (2, 3)$. Putting this into their bound gives

$$\|J^\ell(x, \cdot) - P\|_{\mathrm{TV}} \leq (0.99986)^\ell + (0.998497)^\ell (2 + x).$$

For $x = 0, \theta = 1$, this gives that 34,000 steps are required to make total variation distance smaller than 0.01. Our bounds show that seven steps are required. We are trying to use their ideas for the three component example with joint density $f(j, \theta, n) = \binom{n}{j} \theta^j (1-\theta)^{n-j} \frac{e^{-\lambda} \lambda^n}{n!}$. It does not seem easy.

## 2. TWO MORE FOCUSED RESPONSES

Jones and Johnson suggest that "the existence of a CLT is asymptotic." We disagree. In situations such as the present one, with control on the spectral gap, there are non-asymptotic Berry–Esseen results for additive functions of Markov chains. Useful references are Mann [15] or Lezaud [14]. In situations where one has explicit constants for geometric ergodicity, the work of Kontoyanis and Meyn [11, 12] seems quite explicit.

At the end of their comment Berti et al. suggest that the conditionally reducible families of Consonni and Veronese [5] may be amenable to our explicit analytical techniques. This has recently been pushed through for multinomial, multivariate normal and other examples in Khare and Zhou [10].

## 3. LANCASTER FAMILIES

Gerard Letac has given a masterful summary of this part of our paper along with several new examples. As anyone who studies this subject learns, fascinating new examples "pop out of nowhere." While our paper was being edited, a large new class of processes with polynomial eigenfunctions surfaced in the work of Bryc, Matysiak and Wesolowski [4]. Orthogonal polynomials have a hierarchy; lower ones (e.g., Hermite) being limits of higher ones (e.g., Laguerre, Charlier). At the top of the list are the Askey–Wilson polynomials. These have yet to appear in natural probability problems. Just below them are the very similar Al-Salem Chihara family. These are central to the work of Bryc et al.

Letac rightly points out that our location families are a subclass of models suggested by Eagleson. We would like to point out a strange anthropological feature of this part of the world. In this age of "computational statistic," the kind of distribution theory that Letac (and we) enjoy so much is sometimes regarded as an old fashioned corner of statistics. We recently fielded a question from Susan Holmes who had trivariate count data (668 patients with counts of number of mutations in three regions of each patient's HIV strain). The data had Poisson margins and curious correlations. Because we knew of Eagleson's work and its extensions by Letac [13] and Griffiths et al. [8], along with practical implementation by Karlis and Meligkotsidou [9], we were able to suggest simple models which made good sense (and matched the data). The old fashioned corner shone brightly, at least for a moment. See Rhee et al. [16].

We would like to add one thought to Letac's list of three. We regard one of our major contibutions as the *use* of Lancaster families for explicit determination of rates of convergence of the Gibbs sampler. This allows us to harness years of work by Letac and his students along with the wonderful tools developed by the orthogonal polynomial community to answer simple interesting questions in mathematical statistics.

## 4. SCANNING STRATEGIES

George Casella and Richard Levine bring a fresh perspective, useful literature and great new questions. We continue the discussion in two directions.

### 4.1 Diagonalization for Non-Uniform Scan Strategies

It is not necessary to use uniform coordinate choices to diagonalize our random scan samplers. Suppose that we choose the $x$-coordinate with probability $\alpha$ and the $\theta$ coordinate with probability $1 - \alpha$. Using the setup from Section 3, the corresponding random scan operator $\bar{K}_\alpha$ on $L^2(P)$ is given by

$$\bar{K}_\alpha g(x, \theta) = \alpha \int_\Theta g(x, \theta) \pi(\theta' \mid x) \pi(d\theta') + (1 - \alpha) \int_\mathcal{X} g(x', \theta) f_\theta(x') \mu(dx')$$

For $0 \leq k < c$, consider $\bar{K}_\alpha$ acting on $p_k(x) + uq_k(\theta)$ where $u$ satisfies

(1) $\qquad \alpha u(1 + \mu_k u) = (1-\alpha)(\eta_k + u).$

The result is

$$\alpha(p_k(x) + E_x[q_k(\theta')]u)$$
$$+ (1-\alpha)(E_\theta[p_k(x)] + uq_k(\theta))$$
$$= \alpha(1 + \mu_k u)p_k(x) + (1-\alpha)(\eta_k + u)q_k(\theta)$$
$$= \alpha(1 + \mu_k u)\left[p_k(x) + \frac{(1-\alpha)(\eta_k + u)}{\alpha(1 + \mu_k u)}q_k(\theta)\right]$$
$$= \alpha(1 + \mu_k u)[p_k(x) + uq_k(\theta)].$$

The last equality follows from (1). If $c < \infty$, then for $k \geq c$, Lemma A2 (Appendix) shows that $E_x[q_k(\theta')] = 0$ for all $x$. It follows that $\bar{K}_\alpha$ is diagonalizable with eigenvalues/eigenfunctions

$$\frac{1 \pm \sqrt{(1-2\alpha)^2 + 4\alpha(1-\alpha)\mu_k\eta_k}}{2},$$

$$p_k(x) + \frac{(1-2\alpha) \pm \sqrt{(1-2\alpha)^2 + 4\alpha(1-\alpha)\mu_k\eta_k}}{2\alpha\mu_k}$$
$$\cdot q_k(\theta) \quad \text{for } 0 \leq k < c,$$
$$1 - \alpha, \quad q_k(\theta) \quad \text{for } c \leq k < \infty.$$

In particular, the spectral gap is

$$\frac{1 - \sqrt{(1-2\alpha)^2 + 4\alpha(1-\alpha)\mu_1\eta_1}}{2}.$$

Clearly, the spectral gap is maximized when $\alpha = \frac{1}{2}$. Hence, if we choose spectral gap as a measure of convergence optimality, then uniformly choosing coordinates is the best way. However, as we point out later on, spectral gap is not always the most accurate criterion for measuring convergence optimality, and convergence of Markov chains often depends on more subtle notions.

### 4.2 Comparison with Systematic Scan Strategies

The class of systematic scan strategies is frequently used in practice and just "seems sensible." It is hard to prove things, or compare to random scan strategies, especially for high-dimensional problems, because the systematic scan chains become quite nonlocal. However, for our examples it is not difficult to analyze the random scan chain. We give the details for the beta-binomial case (uniform prior), but mostly everything applies to all of the examples.

Let $\bar{K}$ [defined in (2.3)] be the operator corresponding to the random scan chain. Then $\bar{K} = \frac{1}{2}(P_1 + P_2)$, where

$$P_1 g(x,\theta) = \int_\Theta g(x,\theta')\pi(\theta' \mid x)\pi(d\theta') \quad \forall g \in L^2(P)$$

and

$$P_2 g(x,\theta)$$
$$= \int_\mathcal{X} g(x',\theta) f(x' \mid \theta) m(dx') \quad \forall g \in L^2(P),$$

are the projection operators onto $L^2(m)$ and $L^2(\pi)$, respectively.

PROPOSITION 1. *For the beta/binomial random scan chain (uniform prior),*

$$\|\bar{K}^\ell_{n,\theta} - f\|_{\mathrm{TV}} \geq \frac{1}{3}\left(1 - \frac{1}{n+2}\right)^\ell$$

$$\forall n \geq 1, \theta \geq \frac{1}{2}, \ell \geq 1$$

*and*

$$\|\bar{K}^\ell_{n,\theta} - f\|_{\mathrm{TV}}$$
$$\leq 3e^{-(\ell-1)/8}$$
$$+ 10\sqrt{\frac{n+2}{n}}\left(\frac{1}{2} + \frac{1}{2}\left(1 - \frac{2}{n+2}\right)^{1/2}\right)^{\ell-1}$$

$$\forall n \geq 1, \theta \geq \frac{1}{2}, \ell \geq \frac{3n}{4}.$$

REMARK. Note that $\frac{1}{2} + \frac{1}{2}(1 - \frac{2}{n+2})^{1/2} = 1 - \frac{1}{n+2} + O(\frac{1}{n^2})$. For the systematic scan Gibbs samplers, the distance after $\ell$ steps is roughly (up to small explicit multiplicative constants) $(1 - \frac{2}{n+2})^\ell$. Hence in this sense, the random scan chain takes twice the amount of time as the systematic scan chain to converge to the stationary distribution. Although one might argue that computationally one step of the systematic scan chains is comparable to two steps of the random scan chain, hence they are equivalent computationally.

PROOF OF PROPOSITION 1. The function $\phi(x,\theta) = (x - \frac{n}{2}) + \sqrt{n(n+2)}(\theta - \frac{1}{2})$ is an eigenfunction corresponding to the eigenvalue $\frac{1}{2} + \frac{1}{2}\sqrt{\frac{n}{n+2}}$ for $\bar{K}$. Using the bound of Lemma 2.1 we have

$$\|\bar{K}^\ell_{n,\theta} - g\|_{\mathrm{TV}} \geq \frac{1}{3}\left(\frac{1}{2} + \frac{1}{2}\sqrt{\frac{n}{n+2}}\right)^\ell$$

$$\geq \frac{1}{3}\left(1 - \frac{1}{n+2}\right)^\ell$$

$$\forall n \geq 1, \theta \geq \frac{1}{2}, \ell \geq 1.$$

This shows that $\ell$ must be of order $n$ to have a chance of total variation convergence. We next show that this order of steps suffice.

For the upper bound, we expand $\bar{K}^l$ using the binomial theorem and use the fact that $P_1, P_2$ are projections, so repeated terms cancel out. Thus,

$$\bar{K}^3 = (\tfrac{1}{2}(P_1+P_2))^3$$
$$= \tfrac{1}{8}\{(P_1 + 2P_1P_2 + P_1P_2P_1)$$
$$+ (P_2 + 2P_2P_1 + P_2P_1P_2)\}.$$

Let $K$ and $\widetilde{K}$ be the systematic scan operators defined in Section 2.1. Note that $\widetilde{K} = P_1P_2$ and $K = P_2P_1$. For $\ell \geq \frac{3n}{16}$, it follows from the work done in Section 4 that $\forall n \geq 1, \theta \geq \frac{1}{2}$,

$$\|\widetilde{K}^\ell_{n,\theta} - f\|_{\mathrm{TV}} \leq 10\left(1 - \frac{2}{n+2}\right)^{\ell - 1/2},$$

$$\|K^\ell_{n,\theta} - f\|_{\mathrm{TV}} \leq 10\left(1 - \frac{2}{n+2}\right)^{\ell}.$$

The number of terms in the binomial expansion of $\bar{K}^\ell$ which collapse to $\widetilde{K}^j$ or $K^j$ is easily seen to be $\binom{l-1}{2j-1}$, $1 \leq j \leq \frac{l}{2}$. The number of terms on the binomial expansion which collapse to $\widetilde{K}^j P_1$ or $K^j P_2$ is easily seen to be $\binom{l-1}{2j}, 0 \leq j < \frac{l}{2}$. Note that:

1. $\|\cdot\|_{\mathrm{TV}}$ is convex and $\|\cdot\|_{\mathrm{TV}} \leq 1$.
2. By Azuma's inequality, $\frac{\sum_{j=0}^{l/4}\binom{l}{j}}{2^l} \leq e^{-l/8} \ \forall l \geq 1$.
3. $\|(\widetilde{K}^j P_1)_{n,\theta} - f\|_{\mathrm{TV}} \leq \|\widetilde{K}^j_{n,\theta} - f\|_{\mathrm{TV}}$.
4. $\|(K^j P_2)_{n,\theta} - f\|_{\mathrm{TV}} \leq \|K^j_{n,\theta} - f\|_{\mathrm{TV}}$.

Using the facts above and the binomial expansion of $(\frac{1}{2}(P_1+P_2))^l$, it follows that for $\ell \geq \frac{3n}{4}$,

$$\|\bar{K}^\ell_{n,\theta} - f\|_{\mathrm{TV}}$$
$$\leq 3e^{(\ell-1)/8}$$
$$+ 10\sqrt{\frac{n+2}{n}} \sum_{j=\ell/4+1}^{\ell} \frac{\binom{l-1}{j-1}(\sqrt{n/(n+2)})^{j-1}}{2^{l-1}}$$
$$\leq 3e^{-(\ell-1)/8}$$
$$+ 10\sqrt{\frac{n+2}{n}} \sum_{j=1}^{\ell} \frac{\binom{\ell-1}{j-1}(\sqrt{n/(n+2)})^{j-1}}{2^{l-1}}$$
$$\leq 3e^{(l-1)/8}$$
$$+ 10\sqrt{\frac{n+2}{n}}\left(\frac{1}{2} + \frac{1}{2}\left(1 - \frac{2}{n+2}\right)^{1/2}\right)^{\ell-1}. \quad \square$$

REMARK. The calculations can be carried out for the non-symmetric mixture $\alpha P_1 + (1-\alpha)P_2$. The multipliers of the condensed terms $\widetilde{K}^j, K^j, \widetilde{K}^j P_1$ and $K^j P_2$ are now polynomials in $\alpha$ which can be given explicitly. The asymptotics of these multipliers are available using the distribution theory of the classical Wald–Wolfowitz runs test. We omit further details since in present examples the choice $\alpha = \frac{1}{2}$ seems best.

In the handful of other cases where things can be proved, systematic scan chains are *not* superior to random scan chains. One nice example involves graph coloring. A natural algorithm is to scan through vertices and try a new color. If this results in a legitimate coloring, the change is made, else the previous coloring is kept. Should one choose vertices at random or scan through systematically? Dyer et al. [7] managed to find classes of graphs where random and systematic scans are comparable.

Similar results are found for a natural statistical problem involving a non-uniform distribution on permutations (Mallows model through Kendall's tau). Diaconis and Ram [6] coupled with Benjamini et al. [2] found random pairwise transpositions followed by Metropolis comparable with systematically scanning through all coordinates. The Diaconis and Ram paper contains a literature review of scanning strategies.

A very important point made by Casella and Levine is that asymptotic variance of a few statistics of interest gives an important alternative notion of convergence that can give different answers. This is an important research area. See Bassetti and Diaconis [1] for some first steps/tools.

Finally, we note and mildly object to equating convergence rates with spectral gap. The present authors have spent much of their careers trying to make the point that practical convergence of Markov chains depends on much subtler notions. Consider the Poisson–Gamma example in our Proposition 4.4 with $a = \alpha = 1$. The second eigenvalue of the $x$-chain is $\frac{1}{2}$. If we just use this, we get the usual bound

$$\|K^\ell_j - m\|_{\mathrm{TV}} \leq \sqrt{\frac{1}{m(j)}}\left(\frac{1}{2}\right)^\ell \quad \text{with } m(x) = \left(\frac{1}{2}\right)^{x-1}.$$

This bound implies that it takes $\ell$ of order $j$ steps to randomize. Proposition 4.4 shows that $\ell$ of order $\log j$ steps is the right answer. We may wonder why tuning behaviour to a criterion (like spectral gap) tangentially related to convergence is worthwhile.

Of course, we too have sinned in this direction; see [3].

In conclusion, we thank our discussants and editors for their help, encouragement and good ideas. Thanks to Susan Holmes for the Poisson example and to Guoqiang Hu and Wai Wai Liu for help with the beta-binomial example.